\documentclass{emulateapj}
\slugcomment{{\em}}
\usepackage{epsf}

\begin{document}
\title{The Evolving Faint-End of the Luminosity Function}

\author{S. Khochfar$^1$,  J. Silk$^1$,  R. A. Windhorst$^2$, R. E. Ryan Jr.$^3$ }
 
\affil{$^1$ Department of Physics, Denys Wilkinson Bldg., University
of Oxford, Keble Road, Oxford, OX1 3RH, UK \\
$^2$ School of Earth and Space Exploration, Arizona State University, Tempe, AZ, 85287 \\
$^3$ Department of Physics, Arizona State University, Tempe, AZ, 85287 \\ }

\begin{abstract}
We investigate the evolution of the faint-end slope of the luminosity function, $\alpha$, 
using semi-analytical modeling of galaxy formation. In agreement with observations, we find that 
the slope can be fitted well by $\alpha (z) =a+b z$, with $a=-1.13$ and $b=-0.1$.  The main 
driver for the evolution in $\alpha$ is the evolution in the underlying dark matter mass function.
Sub-$L_*$ galaxies  reside in dark matter halos that occupy a different part of the mass function. 
At high redshifts, this part of the mass function is steeper than at low redshifts, and hence
$\alpha$ is steeper. Supernova feedback in general causes the same relative flattening with 
respect to the dark matter mass function. The faint-end slope at low redshifts is dominated by 
field galaxies and at high redshifts by 
cluster galaxies. The evolution of $\alpha(z)$ in each of these environments is different, with 
field galaxies having a slope $b=-0.14$ and cluster galaxies $b=-0.05$. 
The transition from cluster-dominated  to field-dominated faint-end slope 
occurs roughly at a redshift $z_* \simeq 2$, and suggests that 
a single linear fit to the overall evolution of 
$\alpha(z)$ might not be appropriate. Furthermore, 
this result indicates that tidal disruption of dwarf galaxies in clusters 
cannot play a significant role in explaining the evolution of $\alpha(z)$ at $z< z_*$. 
In addition we find that different star-formation efficiencies $a_*$ in the 
Schmidt-Kennicutt-law 
and supernovae-feedback efficiencies $\epsilon$  generally do not strongly influence the 
evolution of $\alpha(z)$.

\end{abstract}


\section{Introduction}
The galaxy luminosity function (LF) is one of the cornerstones in our understanding of
galaxy formation and evolution. Since the introduction of a fitting function for its
shape by \citet{1976ApJ...203..297S},  the origin of the form of the LF 
function has been a powerful constraint on model building
\citep[e.g.,][]{2003ApJ...599...38B,2007MNRAS.377..285S}. While recent work has focused somewhat on the
luminous end, its evolution with redshift \citep{2007ApJ...654..858B} and the role of dry mergers 
\citep{2003ApJ...597L.117K,2006ApJ...636L..81N}, 
the faint-end provides additional 
 important clues on galaxy formation. Systematic studies of the faint-end slope  in the local universe
reveal differences between  high and low density environments \citep{1998MNRAS.294..193T}, 
as well as for galaxy samples split by morphologies \citep[e.g.][]{1994AJ....108..437M}. The 
underlying physical processes that shape the faint-end of the LF are generally associated with feedback
from supernovae that is effective in heating gas and driving winds in shallow gravitational potentials 
\citep{1986ApJ...303...39D}. Although the implementation of supernova feedback in galaxy 
formation models has been  extensively investigated  \citep[e.g.][]{2003ApJ...599...38B}
for the local galaxy LF, its impact on the redshift evolution on the 
faint-end has not been as well studied. The recent wealth of LF measured to very faint 
magnitudes  in the rest-frame B-band 
\citep[e.g.][]{2003ApJ...592..819B,2003A&A...408..499W,2007ApJ...656...42M,2007astro.ph..3743R} 
and the rest-frame FUV \citep[e.g.][]{2004ApJ...612L..93Y,2005ApJ...619L..15W,2006ApJ...653...53B}
 allows us to test models with high accuracy.

The purpose of this letter is to investigate the underlying driving mechanism for the redshift
evolution of the faint-end slope. Furthermore, we investigate the impact of  supernova feedback
on the rate of star-formation by varying the relevant efficiency parameters.

\section{Model}
In the following we briefly outline our basic modelling approach and refer the 
reader for more details to  \citet{2005MNRAS.359.1379K}, \citet{2006MNRAS.370..902K} 
and reference therein. We generate merger trees for dark matter halos 
using a Monte-Carlo approach based on the extended Press-Schechter formalism 
\citep{1999MNRAS.305....1S}. As we aim to model the faint-end of the LF
 to high redshifts, we need to make sure that the mass resolution 
in our simulations is sufficient. We generate merger trees from dark matter mass 
functions between $z=0$ and $z=6$, and find that resolving each individual merger tree 
down to a mass resolution of $M_{\mbox{min}}=5 \times 10^{9}$ 
M$_{\odot}$ and $M_{\mbox{min}}= 10^{8}$ M$_{\odot}$ at $z \leq 3$ and $z \geq 4$,
 respectively, gives robust results. Once a tree reaches $M_{min}$, we start moving 
the tree forward in time including physical processes 
associated with the baryons within each dark matter halo that include
gas cooling, star-formation, supernova feedback, reionization and merging 
of galaxies on a dynamical friction time-scale. As the focus of 
this letter is on the faint-end of the luminosity function, we will omit 
including prescriptions for AGN-feedback \citep[e.g.][]{2006MNRAS.370..645B}
or environmental effects \citep{env} that mainly influence the bright-end 
of the luminosity function.

The largest impact on the slope at the faint-end comes from  star-formation 
and associated supernova feedback \citep{1986ApJ...303...39D}. 
Faint galaxies generally occupy small
dark matter halos with shallow potential wells, that allow  effective reheating of 
cold gas in the ISM by feedback from supernovae. 
We model star-formation in galaxies using a parametrisation of the global 
Schmidt-Kennicutt-law \citep{1998ApJ...498..541K} according to which 
$\dot{M}_*=a_* M_{cold}/t_{dyn}$, where $a_*$ is a free parameter 
that is indicative of the efficiency 
of star-formation, $M_{cold}$ is  the mass in cold gas in the galactic disk, and
$t_{dyn}$ is the dynamical time-scale of the galaxy. Following the arguments by 
\citet{1986ApJ...303...39D}, we model the amount of cold gas reheated by 
feedback from supernovae with $\dot{M}_{SN}= 4 \epsilon \dot{M}_*  \eta_{SN} 
 E_{SN}/ 3 V_{max}^2 $, with $\epsilon$ as a free parameter that  controls the 
feedback efficiency, $\eta_{SN}$ the number of supernovae per solar mass 
of stars formed, $ E_{SN}= 10^{51}$ erg is the energy released by each supernova,
and $V_{max}$ is the maximum circular velocity of the dark matter halo in 
which the galaxy resides.
 
For each individual galaxy in our simulation, we store the star-formation 
history and generate its $B-$band and FUV rest-frame luminosity function 
at various redshifts using the stellar population models of 
\citet{2003MNRAS.344.1000B}. The faint-end of the luminosity function is then fitted by a simple power law
with slope $\alpha$ as defined in \citet{1976ApJ...203..297S}.  
We fit the faint-end LF at each redshift with a power law spanning a range 
of four magnitudes at z $\leq 3$ and at least two magnitudes at $z> 3$, 
starting at the lowest magnitude $ L_{min}$ that is unaffected by the 
mass resolution of the simulation. We choose this 
approach over fitting the whole LF with a Schechter-fit, because we 
are missing physical effects (see below) in our model that are 
responsible for shaping the bright-end and the knee of the LF.  
In addition, we increase the number of magnitude bins 
and make sure that the fitted values for $\alpha$ are unaffected by the 
bin-size. Fig.1 shows the luminosity function at magnitudes larger than 
the corresponding minimum magnitude for $L_{min}$.
In this study, we simulate a volume
 of $10^6$ Mpc$^3$, which allows us to calculate $\alpha$ robustly up to a 
redshift $z \simeq 6$.

 Throughout this paper, we use the following set of cosmological parameters 
based on the three year Wilkinson Microwave Anisotropy Probe data 
\citep{2007ApJS..170..377S}:
$\Omega_0=0.27$, $\Omega_{\Lambda}=0.73$, $\Omega_b/\Omega_0=0.17$, 
$\sigma_8=0.77$ and $h=0.71$.

\begin{figure}
  \plotone{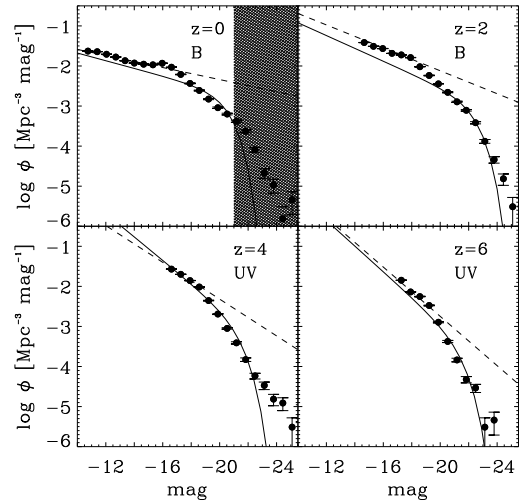}\label{fig4b}
  \figcaption{Comparison between model LFs with 
    $\epsilon=0.6$ and $\alpha=0.02$ (symbols), 
    and Schechter fits to various observations of the LF
    at different redshifts (solid lines). The dashed lines show our power law 
    fits to the faint-end of the LF. 
    Top left panel: Rest-frame $B-$band LF from 
    \citet{2002MNRAS.336..907N}, where we assumed that $B=b_j+0.12$ mag.
    The shaded area indicates the region of the LF
    that is not well matched due to missing feedback effects.  
    Top right panel: Rest-frame $B-$band LF from 
    \citep{2007ApJ...656...42M}. Bottom left: UV-LF 
    from \citet{2007arXiv0707.2080B}. Bottom right: UV-LF
    from \citet{2004ApJ...600L...1Y}. Following \citet{2006ApJ...653...53B}, 
    we assume an average dust correction 0.4 mag at  $z \geq  4$.}
\end{figure}

\section{Results}
There is significant evidence that the faint-end slope of the galaxy luminosity function shows 
a measurable dependence on redshift, which can be fitted by a linear law of the form 
$\alpha (z)= a + b z$, with $a$ between $-1.12$ and $-1.17$, and $b$ between $-0.12$ and $-0.11$ 
\citep[for recent observations, see][]{2006ApJ...648..299S,2007ApJ...656...42M,2007astro.ph..3743R}. 
Within the hierarchical structure 
formation paradigm, one naturally  expects such behaviour, considering that the slope of the 
dark matter mass function below M$_{DM,*}$ is $\alpha_{DM} \sim 2$  and that the objects 
that form in these halos continue to grow by continued star-formation and mergers 
with each other \citep{2001ApJ...561..517K}, hence flattening the slope. 
Fig. 1 shows the predicted model luminosity function at various 
redshifts in the rest-frame $B$ and $UV$. The simulated and observed luminosity 
functions are in fair agreement at the faint-end. The luminous end however, 
shows deviations at low redshift which are due to missing feedback sources in massive galaxies 
such as  AGN or to environmental effects. In Fig. 2, we show the predicted evolution of $\alpha (z) $ for our 
best-fit local model. The free parameters $a_*$ and $\epsilon$ in this model are chosen to give the 
best fit to various local observations \citep[see][]{2006MNRAS.370..902K}. For consistency with 
the majority of observations, we calculate the faint-end slope for the rest-frame FUV at $z \geq 4$ and at 
lower  redshifts for the rest-frame $B$-band. We indeed find an  evolution in $\alpha$ with redshift that is in fair agreement with the observed evolution. 

The immediate question that arises is, what influences and is the main driver for the evolution
in $\alpha$? Generally, supernova feedback is considered the dominant mechanism in shaping the 
faint-end of the luminosity function \citep{1986ApJ...303...39D,2003ApJ...599...38B}. The shaded region 
in Fig. 2 shows the range of linear fits to $\alpha(z)$ that we find by varying the star-formation efficiency
between $a_*=0.02-0.1$ and the supernovae feedback efficiency between $\epsilon=0.2-0.6$. 
 We infer  only a very modest change in $\alpha(z)$ 
for  reasonable choices of feedback efficiencies, and therefore conclude that another process 
must be responsible for the observed evolution in $\alpha(z)$.   

The mass function of dark matter halos is known to show a strong evolution with redshift 
\citep[e.g.][]{1974ApJ...187..425P}. The galaxies contributing to 
the luminosity function around $L_*$ are mostly central galaxies in their dark matter halos, 
i.e. the most luminous galaxy within the halo \citep[e.g.][]{env}. It is therefore not unreasonable 
to assume a connection between the evolution of $\alpha(z)$ and that of the dark matter mass function.
When considering the luminosity of central galaxies residing in dark matter halos of the same mass
at different redshifts, we find that at early 
times, central galaxies are up to three magnitudes brighter than their counterparts in 
low redshift halos. This is even the case 
for halos hosting sub-$L_*$ galaxies. Similar results have been reported by 
\citet{2007MNRAS.376.1465K}, who showed that dwarf galaxies at early times are not affected 
by supernova feedback in their simulations because cooling times are very short in these halos and 
the energy injected by the supernovae is rapidly   dissipated away.  
The slope in the region of dark matter halos that host sub-$L_*$ galaxies is steeper at high
redshift, and  consequently so is $\alpha$. The same is true for other choices 
of $a_*$ and $\epsilon$, thereby  explaining why we do not find any strong dependence of 
$\alpha(z)$ on these parameters. It should be noted however, that modelling these parameters 
with a strong redshift dependency will be able to enhance or weaken the 
evolution of $\alpha$
\begin{figure}
  \plotone{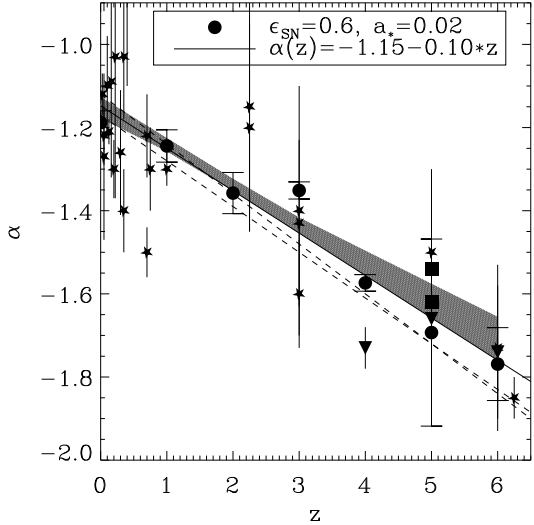}\label{fig1}
  \figcaption{The slope $\alpha$ at different redshifts as 
    predicted by the best fit local model
    with $a_*=0.02$, $\epsilon=0.6$. Filled symbols show results 
    from the simulation and the 
    solid line is the best fit to the simulation data. Errorbars 
    indicate 1-$\sigma$ errors. 
    The dashed lines show the fit to the compiled
    data in \citet{2007astro.ph..3743R}. The shaded region shows
    the range of linear fits to $\alpha(z)$ 
    that we find when varying the star-formation and supernovae 
    feedback efficiencies as discussed in the text. Stars are the 
    compilation from \cite{2007astro.ph..3743R}. The filled squares 
    and triangles show recent results from \citet{2007arXiv0706.2653O} 
    and \citet{2007arXiv0707.2080B}, respectively.}
\end{figure}

We continue analyzing the evolution in $\alpha(z)$ by distinguishing between cluster 
and field galaxies and their relaxation  to the overall luminosity function at the faint-end.
In Fig. 3, we present $\alpha(z)$ for progenitor galaxies of present-day cluster and field galaxies 
from our simulations. 
Here we define cluster environments by present-day dark matter halos above 
$10^{14}$ M$_{\odot}$, and field  environments by halos with masses below 
$10^{12}$ M$_{\odot}$. In field environments 
$ \alpha$ is steeper and evolves more strongly than in cluster environments.  
 At early times, the first galaxies to appear are most likely in high$-\sigma$ 
fluctuations, which will result in present-day galaxy clusters. Consequently, the faint-end 
luminosity function at high redshifts will be dominated by present-day cluster members, and 
the faint-end-slope of the overall galaxy population at high redshifts is flatter than that for the 
field luminosity function alone at the same redshift. 
When considering the relative weight of field galaxies to the overall galaxy population at the 
faint-end, one can estimate the redshift at which the transition from 
cluster-driven to field-driven evolution in $\alpha$ occurs. We find that this
transition roughly occurs at $z_* \simeq 2$ with a slight dependency on the definition of environment.
\begin{figure}
  \plotone{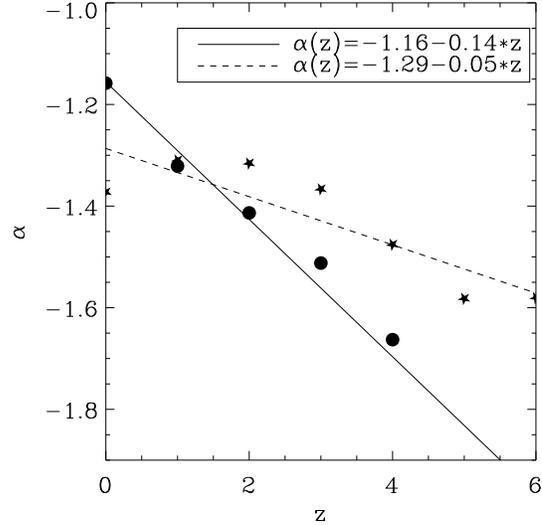}\label{fig4}
  \figcaption{The slope $\alpha$ at different redshifts as predicted by 
    the best fit local 
    model. Filled circles show results for typcal field environments and 
    filled stars show the results for cluster environments, as defined by 
    their present day 
    dark halo mass. The solid and dashed line are the fits to the modeled 
    evolution in the 
    field and cluster, respectively. Here we define cluster environments by 
    present-day dark 
    matter halos above $10^{14}$ M$_{\odot}$ and field  environments, by 
    halos with masses below 
    $10^{12}$ M$_{\odot}$.}
\end{figure}

\section{Discussion and Conclusions}
In this Letter, we presented predictions for the redshift evolution of the faint-end slope of the 
luminosity function within the $\Lambda$CDM-scenario. In general, we find the same trend as
in the recent observations i.e., a steepening  faint-end slope $\alpha$ with redshift, which can be well-fitted by
a simple linear fit $\alpha(z)=a+bz$ where the observations find $a\simeq 1.17$ and $b \simeq -0.11$ 
\citep{2007astro.ph..3743R}. Our simulations predict $a \simeq 1.13$ and $b \simeq -0.1$, in good agreement with the observations considering the large uncertainties, especially 
at high redshifts. 

Our simulations confirm previous results that the flattening of the 
faint-end slope $\alpha$ with respect to the slope in the dark matter mass function 
 can be well explained by supernova feedback. However we additionally show that 
$\alpha$ is steeper at higher redshifts mainly, due to the dark matter mass function being 
steeper for the range of halo masses that host sub-$L_*$ galaxies, suggesting that the evolution 
of $\alpha$ traces closely that of the underlying dark matter mass function.

The contribution of the progenitor population of present-day field and cluster 
galaxies plays a significant role in shaping the evolution  of $\alpha$ .
For field galaxies, the evolution of $\alpha(z)$ is stronger, 
with $b=-0.14$ and  $a=-1.16$ than for cluster galaxies.
In our simulations, we find that at redshifts 
$ z \geq 2$, the faint-end is dominated by galaxies ending up in present-day clusters. 
This transition redshift is dependent on the value of $\sigma_8$, which 
normalizes the power spectrum and regulates the redshift at which structures of a given mass 
typically form. Additionally, the slope of the fluctuation spectrum at  small scales will 
influence $z_*$. Precise high redshift measurements of the contributions of these two populations 
to the faint-end of the luminosity function in future surveys with e.g. the HST WFC3 will help to pin down $z_*$. 
One potential problem for future surveys will be a possible bias towards cluster galaxies as they 
might experience induced star-formation \citep{2007ApJ...654..825M}, increasing their surface 
brightness and making them more easily detectable. This effect will shift $z_*$ to higher redshifts 
and needs to be taken into account carefully.  Observational selection effects will affect the observed 
faint-end LF-slope in Fig. 2. Some observational selection effects \citep[i.e. catalogue incompleteness
and natural confusion,][]{2007astro.ph..3171W} can make the observed faint-end
slope flatter than the true one, while others (e.g., SB-dimming) could make
the observed slope somewhat steeper than the true one, depending on the exact
intrinsic object size distribution. A number of groups correct for
incompleteness either through MC-simulations \citep[e.g.][]{2004ApJ...600L...1Y} or
through cloning techniques \citep[e.g.][]{2006ApJ...653...53B}, and find similar
faint-end slopes when following different procedures. When judging the data,
however, one must keep these observational biases in mind. Ultimately, these
issues can only be resolved with deeper JWST data to AB=31-32 mag.

Tidal disruption of dwarf galaxies in clusters 
as seen in high resolution simulations \citep{1998MNRAS.299..728T}  can in 
principle change the slope $\alpha$. Our results,  however, suggest that at a transition redshift 
of $z_*=2$, the evolution of $\alpha$ changes from
being dominated by cluster galaxies to being dominated by field galaxies. It is therefore not
likely that a large amount of the evolution in $\alpha$ at $z<z_*$ is driven by tidal 
disruption of faint galaxies. An additional implication from the transition at $z_*$ is
that the  evolution of $\alpha(z)$ is better fit by a linear function 
with a break at $z_*$.

The flattening of the slope $\alpha$ with respect to the underlying dark
 matter slope suggests the interesting  possibility of estimating 
the time-scale over which supernovae operate.
Assuming that the first Pop II  stars were formed sometime before 
reionization \citep{2004ApJ...600L...1Y}, and that supernovae type Ia 
originate roughly $ \le 1-2$ Gyr after the bulk of the first Pop II star-formation, 
one would expect an increase in energy  injection into the interstellar 
medium at a redshift corresponding to this time lag.
This additional energy input
will hinder star-formation and contribute to a further increase in the 
mass-to-light ratio of galaxies and hence to an even stronger flattening 
of the slope $\alpha$. It will be crucial to have robust 
measurements of $\alpha$ over a wide range of redshifts 
to probe the onset of the first significant feedback contribution from 
type Ia supernovae. Furthermore, probing the faint-end slope at redshifts 
$ z>6 $, before the significant onset of type II supernovae  will allow us
to measure the underlying dark matter slope very accurately.

Our approach has certain shortcomings. The model presented here did not include any time delay prescriptions for the various SN types, 
but instead assumed instantaneous feedback. More detailed modeling of the time delays and 
its influence on the faint-end slope will be presented elsewhere (in preparation).
Our treatment of supernovae feedback is very simplistic, and more detailed hydrodynamical 
simulations including a multi-phase medium will show if this general trend which we report can be recovered.
First generations of such simulations  indeed show that SN type II that are generated in 
dense star clusters explode into bubbles of hot gas and are therefore less efficient at 
feedback into the ISM \citep{1999ApJ...513..142M} compared 
 to  SN type Ia, which  go off at random places in the galaxy and, can have more effect on the early ISM. \\

We would like to thank Seth Cohen, Evan Scannapieco and Richard Bouwens for helpful discussions and the 
anonymous referee for his useful comments. 
This work was supported by HST grants HST-GO-10530.07 (to RAW) and
HST-AR-10974.01 (to RER) from STScI, which is operated by AURA for NASA under
contract NAS 5-26555, and by NASA JWST grant NAG 5-12460 (to RAW).

\end{document}